\documentclass[twocolumn]{aastex631}
\pdfoutput=1

\hypersetup{colorlinks,linkcolor={blue},citecolor={teal},urlcolor={violet}}

\newcommand{\INFN}{INFN - Sezione di Napoli, Complesso Univ. Monte S. Angelo, I-80126 Napoli, Italy}
\newcommand{\UNI}{Dipartimento di Fisica ``Ettore Pancini'', Università degli studi di Napoli ``Federico II'', Complesso Univ. Monte S. Angelo, I-80126 Napoli, Italy}
\newcommand{\SSM}{Scuola Superiore Meridionale, Università degli studi di Napoli ``Federico II'', Largo San Marcellino 10, 80138 Napoli, Italy}

\shorttitle{Could nearby star-forming galaxies light up the point-like neutrino sky?}
\shortauthors{Ambrosone et al.}

\begin{document}

\title{Could nearby star-forming galaxies light up the point-like neutrino sky?}

\author{Antonio Ambrosone\footnote{ciao}}
\email{aambrosone@na.infn.it}
\affiliation{\UNI}
\affiliation{\INFN}

\author{Marco Chianese}
\affiliation{\UNI}
\affiliation{\INFN}

\author{Damiano F.G. Fiorillo}
\affiliation{\UNI}
\affiliation{\INFN}

\author{Antonio Marinelli}
\affiliation{\INFN}

\author{Gennaro Miele}
\affiliation{\UNI}
\affiliation{\INFN}
\affiliation{\SSM}

\begin{abstract}
Star-forming and starburst galaxies, which are well-known cosmic-rays reservoirs, are expected to emit gamma-rays and neutrinos predominantly via hadronic collisions. In this Letter, we analyze the 10-year Fermi-LAT spectral energy distributions of 13 nearby galaxies by means of a physical model which accounts for high-energy proton transport in starburst nuclei and includes the contribution of primary and secondary electrons. In particular, we test the hypothesis that the observed gamma-ray fluxes are mostly due to star-forming activity, in agreement with the available star formation rates coming from IR and UV observations. Through this observation-based approach, we determine the most-likely neutrino counterpart from star-forming and starburst galaxies and quantitatively assess the ability of current and upcoming neutrino telescopes to detect them as point-like sources. Remarkably, we find that the cores of the Small Magellanic Cloud and the Circinus galaxy are potentially observable by KM3NeT/ARCA with 6 years of observation. Moreover, most of the nearby galaxies are likely to be just a factor of a few below the KM3NeT and IceCube-Gen2 point-like sensitivities. After investigating the prospects for detection of gamma-rays above TeV energies from these sources, we conclude that the joint observations of high-energy neutrinos and gamma-rays with upcoming telescopes will be an objective test for our emission model and may provide compelling evidence of star-forming activity as a tracer of neutrino production.
\end{abstract}


\section{Introduction}

IceCube has been measuring high-energy neutrinos from tens of TeV up to PeVs~\citep{Aartsen:2019fau,Aartsen:2020xpf} during the last decade and yet their origin is still unclear. These measurements certainly reveal the presence of neutrino emitters in our Universe, even though the only clear multi-messenger indication of a neutrino source has been related to the blazar TXS~0506+056 \cite{IceCube:2018dnn}. In this context, star-forming and starburst galaxies (SFGs and SBGs) are abundant sources which could be responsible for at least a part of such observations~\citep{Ambrosone:2020evo}. Indeed, SFGs and SBGs are galaxies endowed with an intense star-forming activity, and therefore they have a particularly high gas density $(n_{\rm ISM} \propto \ 10$--$100\  {\rm cm^{-3}})$, strong magnetic fields $(B\sim 10^2 {\rm \mu G})$, and supernova explosion rates of the order of $0.01$--$1 \ {\rm yr^{-1}}$~\citep{Thompson:2006is}. For these reasons, they are supposed to exhibit strong turbulence which should be able to trap high-energy cosmic-rays inside their environment, giving them an enhanced probability to produce high-energy neutrinos and gamma-rays through hadronic collisions~\citep{Peretti:2018tmo}. In this regard, many authors have studied the properties of the diffuse gamma-ray and neutrino fluxes of these sources~\citep{Loeb:2006tw,Murase:2013rfa,Tamborra:2014xia,Senno:2015tra,Bechtol:2015uqb,Murase:2016gly,Sudoh:2018ana,Peretti:2019vsj,Ajello:2020zna,Ambrosone:2020evo,Blanco:2021icw,Peretti:2021yhc}. Different cosmic-ray transport scenarios have been used to describe the observed gamma-ray emission from SFGs and SBGs, disentangling a calorimetric behaviour from a diffusion-dominated regime~\citep{Peretti:2018tmo,Krumholz:2019uom,Ha:2020nty,Muller:2020vdm,Shimono:2021wvp,Xiang:2021amz,Owen:2021evt,Werhahn:2021jvy,Werhahn:2021bal}. However, in order to explain a sizeable portion of measured high-energy neutrino flux with SFGs and SBGs, the deep Universe must be considered up to redshift $\sim$ 4--5, because of their dimness~\citep{Peretti:2019vsj,Ajello:2020zna,Ambrosone:2020evo}. The low gamma-ray luminosity of SFGs and SBGs typically represents a bound for their contribution to the observed astrophysical neutrinos as a point-like component. Currently, only a dozen of these sources have been catalogued as gamma-ray point-like sources using the Fermi-LAT data and only few of them have been observed through Imaging Cherenkov telescopes.

In this Letter, for the first time we employ a multi-messenger and multi-wavelength approach to assess the ability of current and upcoming neutrino telescopes to observe such galaxies as neutrino point-like sources. This is a question of utmost importance after the IceCube collaboration reported a $2.9\sigma$ excess of neutrino events coming from the direction of NGC~1068 analyzing $10$-year data~\citep{Aartsen:2019fau}. Even though additional Active Galactic Nuclei (AGN) activity could be present, observations like this one suggest that star-forming activity could trace the hadronic emission and produce point-like excesses in the TeV sky observed by IceCube as well as in the future skymap of the upcoming KM3NeT/ARCA and IceCube-Gen2 telescopes.

In this work, following the scenario proposed by \cite{Peretti:2018tmo} which considers the transport of high-energy protons and electrons, we analyze the 10-year Fermi-LAT data provided by~\cite{Ajello:2020zna}, taking into account the spectral energy distributions (SEDs) of 13 SFGs and SBGs. We describe these gamma-ray observations through the hadronic and leptonic processes at work in the star-forming regions and determine the corresponding neutrino fluxes. For each SBG, we require the star formation rate (SFR) to be consistent with the one derived from infra-red (IR) and ultra-violet (UV) observations~\citep{Kornecki:2020riv} within a maximal level of discrepancy, thus making our predictions more robust. Hence, we compare the most-likely neutrino flux normalizations at $1 \ \text{TeV}$ with IceCube, IceCube-Gen2, and KM3NeT sensitivities (see Figure~\ref{fig:neutrino}). Besides, for the brightest sources we also show in Figure~\ref{fig:gamma} the prospects of the Cherenkov Telescope Array (CTA).

\section{Modelling the neutrino and gamma-ray emission}

In this section, we summarize the emission model adopted for the neutrino and gamma-ray spectra of SFGs and SBGs. We adapt the model provided by~\cite{Peretti:2018tmo} (see also~\cite{Peretti:2019vsj,Ambrosone:2020evo} for further details). We assume that the star-forming activity is limited to a small spherical region of the galaxy called starburst nucleus (SBN). Hence, we solve the leaky-box equations to determine the energy distribution of high-energy cosmic-rays (protons and electrons) injected by supernovae explosions. The solution of the leaky-box equations can be written as
\begin{equation}\label{leakybox}
F_{p,e} = Q_{p,e}\left(\frac{1}{T_{\text{adv}}} + \frac{1}{T_{\text{loss}}}+ \frac{1}{T_{\text{diff}}}\right)^{-1} \,,
\end{equation}
where $F_{p,e}$ and $Q_{p,e}$ are respectively the distribution function and the injection rate for protons and electrons. The cosmic-ray distribution stems from a balance of different processes with characteristic timescales: the advection timescale $T_{\text{adv}} =  R_\mathrm{SBN}/v_{\text{wind}}$, with $R_\mathrm{SBN}$ being the SBN radius and $v_{\text{wind}}$ the dust-wind velocity, the energy losses timescale $T_{\text{loss}}$, and the diffusion timescale $T_{\text{diff}}$. For protons,  energy losses are dominated by ionization, proton-proton, and Coulomb interactions, whereas for electrons they are dominated by ionization, Inverse Compton scatterings, and synchrotron interactions~\citep{Peretti:2018tmo}. For the cosmic-ray diffusion, we consider a Kolmogorov-like model with diffusion coefficient $D(E) \propto E^{1/3}$, so that $T_{\text{diff}}(E) \propto E^{-1/3}$. The diffusion timescale is typically larger than the other ones due to the large magnetic fields and the high levels of turbulence of SBGs and SFGs. This implies that the gamma-ray and neutrino emission is only marginally affected by the specific choice of the diffusion model.

We assume that protons are injected with a power-law spectrum in momentum space with spectral index $\Gamma+2$ and an exponential cutoff at $10~\mathrm{PeV}$, in agreement with the combined fit of IceCube and Fermi-LAT diffuse data~\citep{Ambrosone:2020evo}. The precise value of the cutoff energy is irrelevant for our purposes, since the gamma-ray flux is suppressed already at much smaller energies by internal and external absorption. The proton spectrum is directly proportional to the star formation rate $\dot{M}_*$ and normalized by requiring that each supernova releases into protons $10\%$ of its total explosion kinetic energy ($\sim 10^{51}~\mathrm{erg}$). On the other hand, electrons have the same spectral index, a Gaussian cutoff at $10~\mathrm{TeV}$, and a normalization fixed to one fiftieth of the one of the protons, similar to what is inferred for our Galaxy \citep{Torres:2004ui,Peretti:2018tmo}.

Neutrinos are emitted through the decay of charged pions ($\pi \to \mu \,\nu_\mu$, $\mu \to e\,\nu_e\,\nu_\mu$) that are produced in hadronic interactions of the injected protons with the interstellar gas. We determine the neutrino production rate by assuming that pions always carry 17\% of their parent proton energy~\citep{Kelner:2006tc}. Gamma-rays are emitted in hadronic processes through neutral pion decays ($\pi^0\to \gamma\gamma$) and in leptonic processes through bremsstrahlung and Inverse Compton scatterings of electrons. We also take into account the contribution of secondary electrons produced by the decay of charged pions. We emphasize that both the hadronic and leptonic components examined in this study are dictated by the star formation rate. Investigating the possible presence of additional components with a different origin (e.g. related to AGN activities) is beyond the scope of this work. The gamma-ray emission from Inverse Compton scatterings depends on the density of background photons acting as a target. In~\cite{Peretti:2018tmo}, such a radiation is modeled as a superposition of black-body spectra. However, as emphasized also there, we can approximate it as a monochromatic spectrum with an energy equal to the peak energy $\epsilon_{\rm peak} = 0.1 \ \text{eV}$. Therefore, the Inverse Compton spectrum can be simply characterized by the energy density $U_\mathrm{rad}$ of the background photons. Analogously to~\cite{Ambrosone:2020evo}, we consider all sources to have a similar spectral shape for the background photons equal to the M82 best-fit background spectrum reported by~\cite{Peretti:2018tmo}. The normalization for the different sources is self-consistently determined by the radiation energy density of each source $U_{\text{rad}}$ (see below). Finally, we account for internal and external gamma-ray absorption due to pair-production processes with background photons. For the latter, we consider as a target CMB photons as well as the extragalactic background light model reported in~\cite{Franceschini:2017iwq}. We point out that gamma-ray absorption is relevant at energies larger than a few TeV, and therefore it affects only the forecast predictions for the CTA telescope.

The neutrino and gamma-ray emission mainly depends on the spectral index $\Gamma$ and on the star formation rate. For the density $n_\mathrm{ISM}$ of the interstellar gas (target of the proton interactions), we rely on the Kennicutt relation~\citep{Kennicutt:1998zb,Kennicutt:2012ea,Kennicutt_2021}. This relation connects the surface density of SFR, $\Sigma_{\rm SFR} = \dot{M}_*/(\pi R_\mathrm{SBN}^2)$, and the gas surface density, $\Sigma_{\text{gas}} = m_p\,n_\mathrm{ISM}\,R_\mathrm{SBN}$. For simplicity, we consider for all the galaxies $R_\mathrm{SBN}=200~\mathrm{pc}$, which is an average size of the circumnuclear regions~\citep{Peretti:2018tmo}. By using the benchmark value $n_\mathrm{ISM} = 175~{\rm cm^{-3}}$ for $\dot{M}_*= 5~{\rm M_\odot\, yr^{-1}}$~\citep{Peretti:2018tmo}, we obtain the scaling relation
\begin{equation}\label{eq:kennicutscaled}
    n_{\rm ISM} = 175 \left(\frac{\dot{M}_*}{5~{\rm M_\odot\, yr^{-1}}}\right)^{2/3}\,{\rm cm^{-3}} \,.
\end{equation}
For the background photon density $U_{\rm rad}$, we assume a direct proportionality to the SFR, which is expected to be tightly related to the infrared (IR) luminosity~\citep{Kennicutt:1998zb,Inoue:2000hm,Hirashita:2003su,2011PASJ...63.1207Y,Kennicutt:2012ea,Kennicutt_2021}. In particular, we consider the scaling relation
\begin{equation}\label{eq:infrared}
    U_\mathrm{rad} = 2500 \left(\frac{\dot{M}_*}{5 \, {\rm M_\odot\, yr^{-1}}}\right)\,{\rm eV \,cm^{-3}} \,,
\end{equation}
where the reference values, $U_\mathrm{rad} = 2500 \,{\rm eV \,cm^{-3}}$ and $\dot{M}_* = 5\,{\rm M_\odot\, yr^{-1}}$, are obtained from what has been inferred for M82 by~\cite{Peretti:2018tmo,Peretti:2019vsj}. Such a relation is generally satisfied by high-SFR galaxies which behave as good calorimeters. On the other hand, low-SFR galaxies have been shown to deviate from the assumed linear SFR-IR relation, as also recently pointed out by~\cite{Kornecki:2020riv,Werhahn:2021bal}. Nevertheless, for these galaxies we have verified that equation~\ref{eq:infrared} is a good and conservative approximation: indeed the gamma-ray emission depends only slightly on $U_\mathrm{rad}$ since the leptonic component is in general subdominant in our model. Moreover, a smaller value for $U_{\rm rad}$, as observed for low-SFR galaxies~\citep{Kornecki:2020riv}, would further reduce the leptonic production in favor of the hadronic one. Finally, for all the galaxies, we adopt the average values $B = 200~\mu{\rm G}$ \citep{Thompson:2006is} and $v_{\rm wind}=500~{\rm km/s}$~\citep{Peretti:2019vsj} for the magnetic field and the dust-wind velocity, respectively. 

\begin{deluxetable*}{lccccc}
\tablecaption{Results of the likelihood analysis of current gamma-ray data. The columns report the source name, the SFR prior, the most-likely values of the two parameters, the 68\% maximum posterior density credible intervals of the marginal distributions, and the reduced chi-squared values considered as an estimate of the goodness of the fit.
\label{tab:tab}}
\tablehead{\colhead{Source} & \colhead{Uniform prior} & \colhead{Most-likely values} & \multicolumn{2}{c}{68\% credible intervals} & \colhead{$\chi^2 / \mathrm{dof}$} \\
& \colhead{$\dot{M}_*$} & \colhead{$(\dot{M}_*,\,\Gamma)$} & \colhead{$\dot{M}_*$} & \colhead{$\Gamma$}  & \colhead{}}
\startdata
M82 & 3.0 -- 30 & $\left(4.5 , \, 2.30\right)$ & $\left[4.3, \, 4.6\right]$ & $\left[2.27,\, 2.33\right]$ & 1.24 \\
NGC 253 & 1.4 -- 17 & $\left(3.3, \, 2.30\right)$ & $\left[3.14, \, 3.40 \right] $ & $\left[ 2.28 ,\, 2.32 \right] $ &  1.32 \\
ARP 220  & 60 -- 740 & $\left( 740 ,\, 2.66 \right)$ & $\left[ 492,\, 740 \right]$ & $\left[ 2.51,\, 2.68 \right] $ & 1.52 \\
NGC 4945  & 0.35 -- 4.15 & $\left( 4.15,\, 2.30 \right)$ & $\left[ 4.05,\, 4.15 \right]$ & $\left[ 2.23,\, 2.32 \right] $ & 1.52 \\
NGC 1068  & 5 -- 93 & $\left( 16,\, 2.52\right)$ & $\left[ 13,\,20 \right] $ & $\left[ 2.45,\, 2.65 \right] $ & 0.65 \\
NGC 2146& 3 -- 57 & $\left( 15,\, 2.50 \right)$ & $\left[ 9,\, 27 \right] $ & $\left[ 2.44,\, 2.88 \right]$ & 0.50\\
ARP 299 & 28 -- 333 & $\left( 28,\, 2.15\right)$ & $\left[ 28,\, 200 \right]$ & $\left[ 1.40 ,\, 1.90 \right] \cup \left[  2.77,\, 3.00 \right] $ & 0.18 \\
M31 & 0.09 -- 0.90 & $\left( 0.34 ,\, 2.40 \right)$ & $\left[ 0.31 ,\, 0.40 \right] $ & $\left[ 2.29 ,\, 2.61 \right] $ & 0.52 \\
M33 & 0.09 -- 0.90 & $\left( 0.44 ,\, 2.76 \right)$ & $\left[ 0.19 ,\, 0.56 \right] $ & $\left[ 2.57 ,\, 2.96 \right] $ & 0.44 \\
NGC 3424 & 0.4 -- 5.4 & $\left( 5.4,\, 2.22 \right)$ & $\left[ 2.5,\, 5.4 \right]$ & $\left[ 1.92 ,\, 2.67 \right] $ & 1.63 \\
NGC 2403 & 0.1 -- 1.2 & $\left( 0.75 ,\, 2.12 \right)$ & $\left[ 0.58 ,\, 0.96 \right] $ & $\left[ 1.92 ,\, 2.36 \right] $ & 0.38 \\
SMC & 0.008 -- 0.090  & $\left( 0.038,\, 2.14 \right)$ & $\left[ 0.037,\, 0.039 \right] $ & $\left[ 2.13 ,\, 2.16 \right] $ & 1.90\\
Circinus Galaxy& 0.1 -- 8.1 & $\left( 6.6 ,\, 2.32 \right)$ & $\left[ 6.2 ,\, 7.8  \right] $ & $\left[ 2.15 ,\, 2.45 \right] $ & 0.92 \\
\enddata
\tablecomments{The star formation rate $\dot{M}_*$ is in units of $\text{M}_\odot \,\text{yr}^{-1}$.}
\end{deluxetable*}
\section{Analysis and Results}

Our main purpose is to determine the high-energy neutrino emission from SFGs and SBGs based on a likelihood analysis of gamma-ray data. A crucial observation must be made: for some of these galaxies, a possible additional source of gamma-rays may be related to AGN activity~\citep{Alonso-Herrero:2013mpa,Yoast-Hull:2017wwl,Guo:2019mpj,Inoue:2019yfs,Murase:2019vdl,Ajello:2020zna,Xiang:2021jow,Kheirandish:2021wkm}, which is not included in the following analysis. Thus we test the hypothesis that the observed gamma-ray SEDs are saturated by star-forming activity only, and determine the most-likely neutrino emission under this assumption. We analyze the gamma-ray SEDs of 13 galaxies observed by Fermi-LAT with 10 years of observation~\citep{Ajello:2020zna}.\footnote{The data reported in~\cite{Ajello:2020zna} are normalized to a constant value of $10^{-9} \, \text{GeV} \, \text{cm}^{-2} \, \text{s}^{-1}$ at an energy of $1$~GeV. We have suitably rescaled them using the reported best-fit parameters of the power-law model.} For M82 and NGC~253 we make also use of the data provided by VERITAS~\citep{2009Natur.462..770V} and H.E.S.S.~\citep{2018A&A...617A..73H}, respectively. For each galaxy, we pursue a Bayesian approach to assess the most-likely values for the two free parameters of the model: the star formation rate $\dot{M}_*$ and the spectral index $\Gamma$ of injected protons and electrons. We determine the posterior distribution as
\begin{equation}\label{eq:post}
    p(\dot{M}_*,\,\Gamma|{\rm SED}) \propto p({\rm SED} | \dot{M}_*,\,\Gamma)\, p(\dot{M}_*) \, p(\Gamma) \,,
\end{equation}
with a Gaussian likelihood function
\begin{equation}\label{eq:like}
    p({\rm SED} | \dot{M}_*,\,\Gamma) = e^{-\frac{1}{2}\sum_{i}\left(\frac{\text{SED}_{i}-E^2_{i}\Phi_{\gamma}(E_i| \dot{M}_*,\,\Gamma)}{\sigma_{i}}\right)^2} \,.
\end{equation}
Here, $\text{SED}_{i}$ are the measured data, where $i$ runs over the energy bins centered around the energy $E_{i}$, and $\sigma_{i}$ are the observational uncertainties. The data are compared to the gamma-ray emission $\Phi_{\gamma}(E_i| \dot{M}_*,\,\Gamma)$ predicted by our model. We adopt the source distances provided by \cite{Kornecki:2020riv} and neglect the uncertainty on such values. For all the galaxies we consider the same uniform prior on the spectral index in the range 1.0--3.0. For the SFR, we instead adopt a different uniform prior distribution for each source. We require $\dot{M}_*$ to be consistent within a factor of 3 with the corresponding values reported in~\cite{Kornecki:2020riv}. This choice is made to cover the wide variety of SFR estimates presented in the literature~\citep{Groves_2008,Bolatto_2011,For_2012,10.1093/mnras/stv2951,Yoast-Hull:2017wwl,Peretti:2019vsj,Ajello:2020zna}.

The results of the analysis are summarized in Table~\ref{tab:tab}. For M82, NGC 253, NGC~2146, NGC~2403, Circinus galaxy, M31, M33, and SMC, we find a reasonable agreement both with the data and the measured SFR values. We emphasize that in our model all the galaxies are approximated as point-like sources with a dominant emission limited to the SBN. Hence, the SBN extension intrinsically defines the extension of the gamma-ray source emission. Most of the sources have not been resolved and therefore they are consistent with our starburst nucleus approximation. On the other hand, SMC and M31 have been observed by \cite{Ajello:2020zna} as extended sources with an angular extension of $1.5^{\circ}$ and $0.45^{\circ}$, respectively. Thus, for these sources we have also performed our analysis by assuming a different core radius in agreement with these observations (i.e. $800$ pc and $3000$ pc, respectively) and have obtained results similar to the ones reported in Fig.~\ref{fig:neutrino}. Indeed, the most likely values of the neutrino flux normalizations change less than a factor of two, thereby making our conclusions unchanged. We stress that our main goal is to assess the possibility for future neutrino telescope of singularly observing these sources, only using their gamma-ray data and the star formation rate as constraints for their neutrino emissions. In order to robustly constrain the parameters of these extended sources, it would be necessary a numerical simulation of their astrophysical environment, which we leave for future work.

For NGC~4945, NGC~3424, and ARP~220, we find a lower limit for the SFR, implying the need for additional contributions to the gamma-ray production. Indeed these sources host an AGN~\citep{Yoast-Hull:2017wwl,Ajello:2020zna,Xiang:2021jow}. A special case is ARP~299, for which we find an upper limit for the SFR. Most of the gamma-ray measurements of this source are only flux upper limits, thus jeopardizing the likelihood procedure. Finally, for NGC 1068, we find a good agreement both with the data and the SFR value given by~\cite{Kornecki:2020riv}. However, as will be discussed later, the corresponding neutrino emission predicted by our model is much lower than the one released by IceCube as a $2.9\sigma$ excess~\citep{Aartsen:2019fau}. Therefore, while the gamma-ray flux could well be explained by the star-forming activity of the source, the explanation of its potential neutrino emission might require an intense AGN hot corona activity~\citep{Inoue:2019yfs,Ajello:2020zna,Kheirandish:2021wkm,Anchordoqui:2021vms}.

\section{Prospects for multi-messenger observations}

\begin{figure}[t!]
    \centering
    \includegraphics[width=\columnwidth]{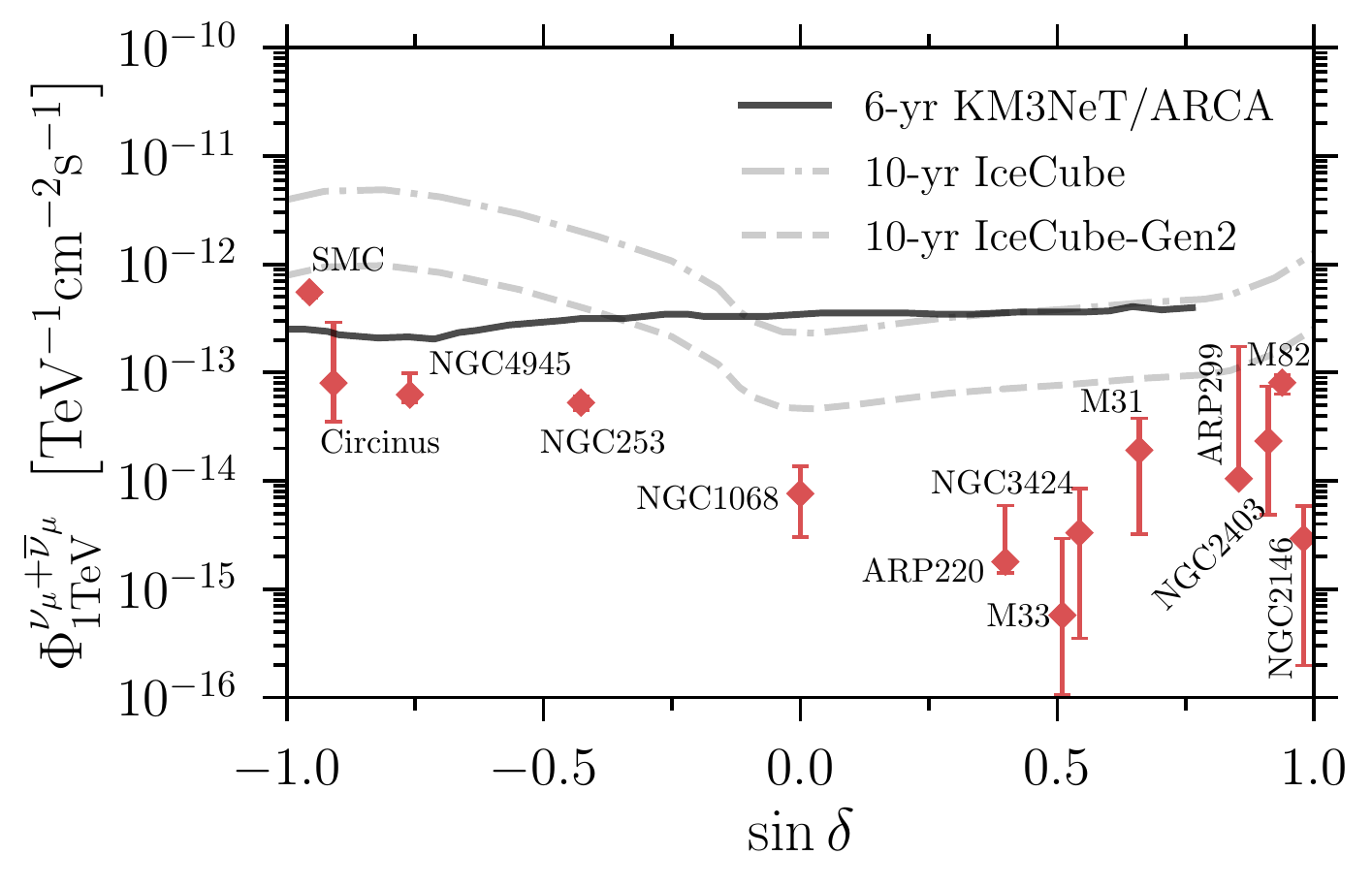}
    \caption{Muon neutrino flux normalizations at $1~\mathrm{TeV}$ predicted by the source star-forming activity as a function of source declination. The diamonds correspond to the most-likely values of the source parameters deduced by current gamma-ray data (see Table~\ref{tab:tab}), while the bands represent the 68\% credible intervals of the marginal flux distributions. The lines shown the point-like sensitivity of different neutrino telescopes: 6-year KM3NeT/ARCA~\citep{Aiello:2018usb}, 10-year IceCube~\citep{Aartsen:2019fau}, and 10-year IceCube-Gen2 estimated according to~\cite{Aartsen:2020fgd}.}
    \label{fig:neutrino}
\end{figure}

The results of the likelihood analysis of current gamma-ray data are hence employed to estimate the high-energy muon neutrino flux from each source. The normalization of such fluxes at $1~\text{TeV}$ are shown in Figure~\ref{fig:neutrino}, where the diamonds represent the predictions for the most-likely source parameters and the bands cover the 68\% interval of the marginal flux distributions. In the plot, our predictions are compared with the IceCube point-like sensitivity~\citep{Aartsen:2019fau}, as well as with the ones of the upcoming neutrino telescopes KM3NeT/ARCA~\citep{Aiello:2018usb} and IceCube-Gen2 with different observation times. The latter is estimated to be at least five times better than the current IceCube one~\citep{Aartsen:2020fgd}. Remarkably, we find that the cores of SMC and Circinus galaxy could be potentially detected by KM3NeT thanks to its higher sensitivity to the southern sky. In the northern sky, a promising source is ARP~299 which might be within reach of IceCube-Gen2. Other galaxies such as NGC~4945, M31, NGC~2403 and M82, are instead just a factor of $\sim 3$ below the KM3NeT and IceCube-Gen2 point-like sensitivities within their 68\% credible intervals. These considerations are likely to be representative of the real detection potential for these sources. It is worth noticing that the experimental sensitivities are determined for an $E^{-2}$ neutrino spectrum, and therefore the conclusions might change for softer spectra. However, as reported in Table~\ref{tab:tab}, the most-likely spectral index for most galaxies is close to 2. Moreover, the largest values for the neutrino flux normalization reported in Figure~\ref{fig:neutrino} are obtained in case of hard spectra, thus supporting the comparison with the reported point-like sensitivities.

\begin{figure*}[t!]
    \centering
    \includegraphics[width=0.48\textwidth]{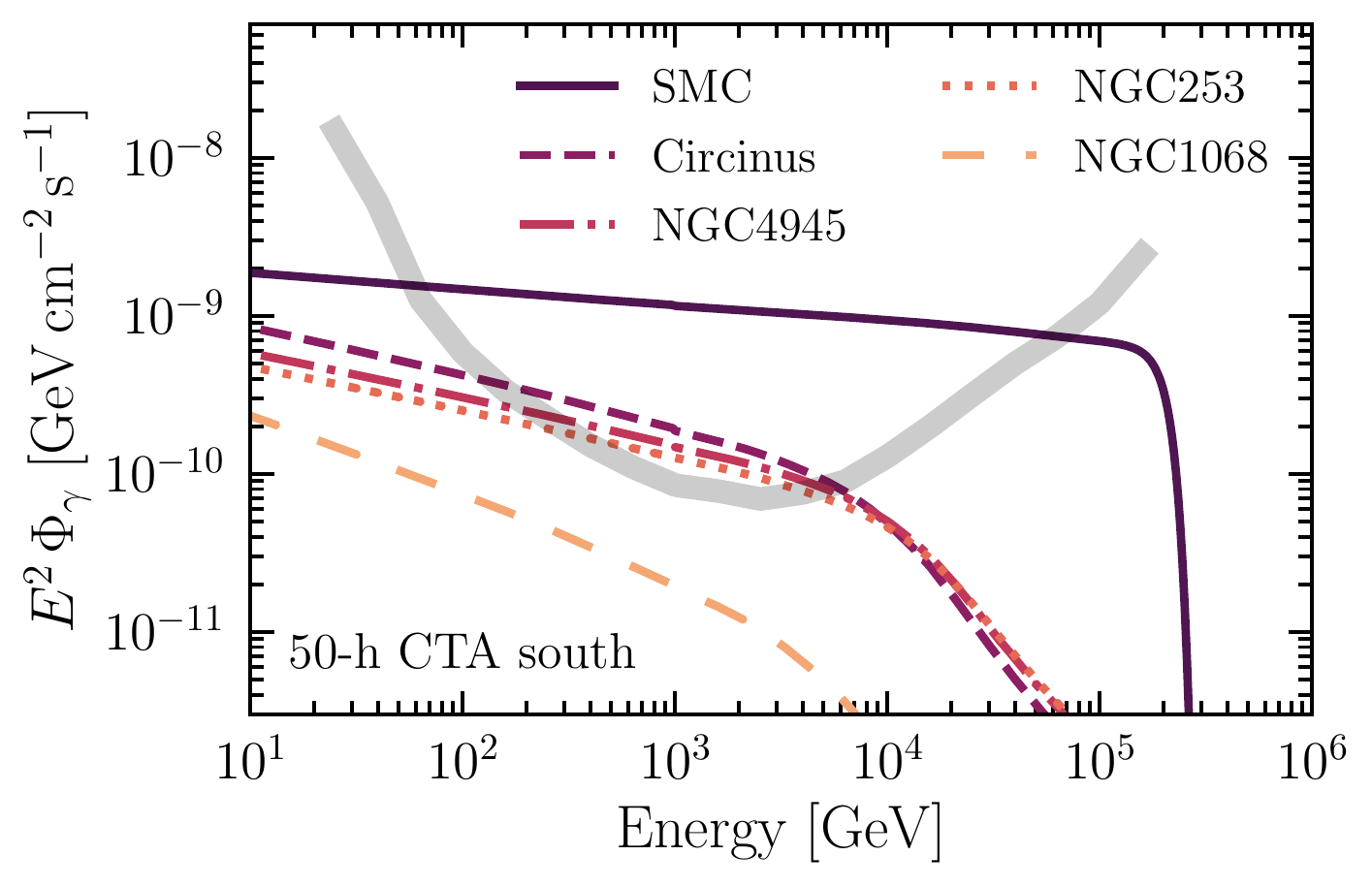}
    \hspace{0.02\textwidth}
    \includegraphics[width=0.48\textwidth]{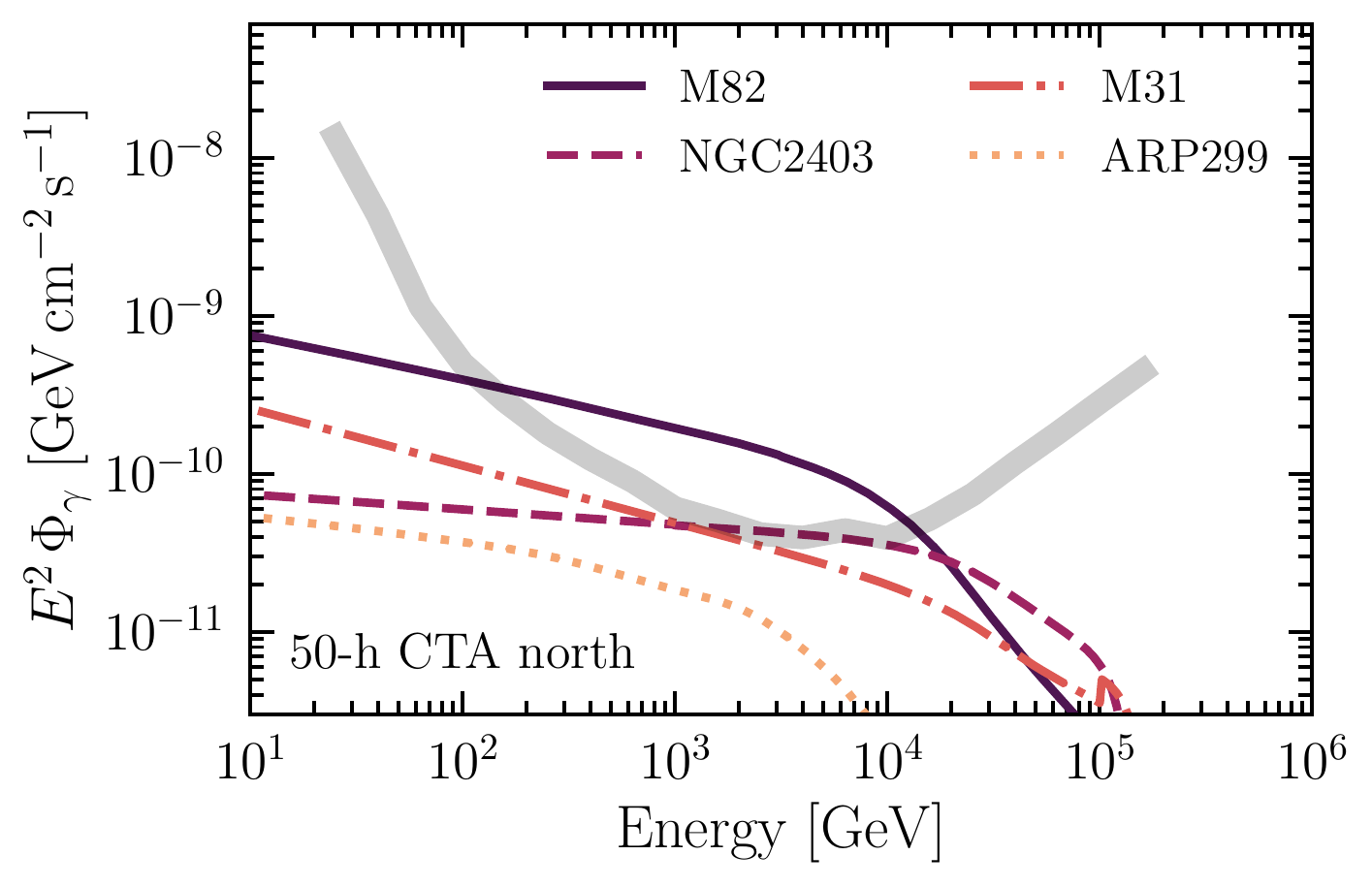}
    \caption{Most-likely gamma-ray spectral energy distributions of different galaxies as a function of gamma-ray energy, in agreement with their star-forming activities and current gamma-ray observations. The left (right) panel collects the galaxies of the southern (northern) sky. The thick lines show the CTA differential sensitivity for an observation time of 50 hours~\citep{CTAConsortium:2018tzg}.}
    \label{fig:gamma}
\end{figure*}

For the brightest neutrino galaxies, we also show in Figure~\ref{fig:gamma} the most-likely very-high-energy gamma-ray flux to be compared with the CTA differential sensitivity, for which we consider an observation time of 50~hours towards the direction of the galaxy~\citep{CTAConsortium:2018tzg}. The cutoff of the gamma-ray flux is due both to internal and external absorption. A few sources are expected to be above the CTA sensitivity in the range between $100$ GeV and $10$ TeV; for SMC this range extends to $100$ TeV. This is particularly relevant, since one of the main limitations in our study is the lack of gamma-ray measurements in the range above 1 TeV (apart from the VERITAS and H.E.S.S. data on M82 and NGC~253). Therefore, our results suggest that CTA will allow us to draw more robust conclusions on the gamma-ray production in SFGs and SBGs. It is worth emphasizing that, under the assumption of a dominant hadronic production linked to the star-forming activity, at high energies the gamma-ray and neutrino emission are directly related each other through the SFR which is derived from the UV and IR observations.

\section{Conclusions}

In the near future, upcoming neutrino telescopes will potentially observe nearby star-forming and starburst galaxies as point-like sources. In particular, they could pose a solid link between the hadronic emission of these sources, supposed to dominate the GeV-TeV gamma-rays, and the expected intense star-forming activity as obtained from IR and UV observations.

As the brightest sources predicted by our emission model are in the southern sky, a leading role will be played by neutrino telescopes located in the northern hemisphere. They include KM3NeT/ARCA as well as the planned Baikal-GVD~\citep{Avrorin:2019dli} and P-ONE~\citep{Agostini:2020aar} telescopes. On the other hand, IceCube-Gen2 will have a better point-like sensitivity in the northern sky, thus remaining a crucial observer for the sources positioned in this portion of the sky. In any case, the advent of a Global Neutrino Network would be necessary to observe most of the point-like emission from SFGs and SBGs predicted in this work, by increasing the available worldwide effective volume of neutrino telescopes.

While our results are mainly derived from the star-forming activity, we cannot exclude a possible AGN emission counterpart for some of the galaxies selected. However, an additional AGN activity would typically exhibit a flaring behaviour, unless the related duty cycle is very high. Therefore, the neutrino measurements over large observational periods would be dominated by the star-forming steady component, thereby making our detection prospects more robust. In this regard, a crucial role will be also played by CTA that will potentially measure the gamma-ray emission above tens of TeV energies for some of the galaxies investigated. This will allow us to place better constraints on the sources production mechanism as well as on the parameters of the emitting cores.

This work indicates that the next decade will be decisive to assess whether the local star-forming activity can be a tracer of point-like neutrino production, and it highlights the importance of pivoting the presented scenarios through the low-energy thermal emissions.

\section*{Acknowledgements}
This work was partially supported by the research grant number 2017W4HA7S ``NAT-NET: Neutrino and Astroparticle Theory Network'' under the program PRIN 2017 funded by the Italian Ministero dell'Universit\`a e della Ricerca (MUR). The authors also acknowledge the support by the research project TAsP (Theoretical Astroparticle Physics) funded by the Istituto Nazionale di Fisica Nucleare (INFN).

\bibliography{biblio}{}
\bibliographystyle{aasjournal}
\end{document}